\documentclass[aps,prl,twocolumn,amsmath]{revtex4}
\usepackage{bm}
\usepackage{graphicx}
\begin{document}
\setlength{\arraycolsep}{2pt}

\title{Reply to "Comment on Phys. Rev. Lett. 105, 170404 (2010)"}
\author{Se-Wan Ji$^{1,2}$, Jaewan Kim$^{2}$, Hai-Woong Lee$^{3}$, M. S. Zubairy$^{4}$, and Hyunchul Nha$^{1,2,*}$} 
\affiliation{$^1$Department of Physics, Texas A \& M University at Qatar, Doha, Qatar\\
$^2$School of Computational Sciences, Korea Institute for Advanced Study, Seoul 130-012, Korea\\
$^3$Department of Physics, Korea Advanced Institute of Science and
Technology, Daejeon 305-701, Korea\\
$^4$Department of Physics and Institute of Quantum Studies, Texas A\& M University, College Station, TX 77843, USA}
\maketitle

In the preceding Comment \cite{Daniel}, Cavalcanti and Scarani (CS) raised a criticism against the proposed Bell test for continuous variables in our Letter \cite{Nha}.
They argue that our inequalities do not make a strong form of Bell inequalities as these rely on another assumption about physical implementation beyond the conditions of locality and realism.
Indeed, we used a physical constraint, $X_j^2+Y_j^2=N_j$, where $X_j$ and $Y_j$ are viewed as the real and the imaginary components of a complex amplitude and $N_j$ its intensity at local systems $j=1,2$ in our inequalities, as also mentioned in the footnote [22] of our Letter \cite{Nha}.
In contrast, if one relies only on the conditions of locality and realism with no reference to physical situations, some counter-models may be constructed to account for the violation of our inequalities like the one
\begin{eqnarray}
\left(\langle X_1X_2\rangle+\langle Y_1Y_2\rangle\right)^2+\left(\langle X_1Y_2\rangle-\langle Y_1X_2\rangle\right)^2\le\langle N_1N_2\rangle,\nonumber
\end{eqnarray}
e.g. the model $X_j=Y_j=1$ and $N_j=0$ ($j=1,2$) by CS \cite{Daniel}.
In this view, our inequalities may not address nonlocality in the strongest sense as CS argue.
However, they may be regarded as the Bell inequalities for testing those local realistic theories that
admit the constraint $X^2+Y^2=N$.

Our reasoning used to derive inequalities in \cite{Nha}, without resorting to quantum mechanics, is that if a single realistic object possesses a nonzero amplitude ($X_j\ne0$ or $Y_j\ne0$), it must possess a nonzero intensity as well ($N_j\ne0$).
This may hold for a good-will scenario where no malicious party intervenes the Bell test. On the other hand, CS envision a situation where
a malicious third party manipulates measuring devices such that $X$ and $Y$ are measured for $H$-polarized light and $N$ for $V$-polarized light that can realize their model
$X_j=Y_j=1$ and $N_j=0$. The logic enabling this counter-model or its variants may be that $N$ can be regarded as a variable completely independent of  $X$ and $Y$ in our inequalities.
To argue against it, one may attempt to experimentally demonstrate the local relation among the three variables $X_j,Y_j$, and $N_j$, which can be tested on an ensemble-level, in addition to the correlation between systems.

On the other hand, one may obtain strong Bell inequalities without additional assumptions from Eqs. (3) and (4) of our Letter by addressing the intensity of the complex amplitude $C_j=C_{jx}+iC_{jy}$ in terms of  the two observables ${\hat C}_{jx}$ and ${\hat C}_{jy}$ only. This approach leads to the inequalities (5) and (6) of  \cite{Nha} with the right-hand terms replaced by $\langle \prod_{j=1,2}\left(\hat{C}_{jx}^2+\hat{C}_{jy}^2\right)\rangle$ and $\prod_{j=1,2}\langle \hat{C}_{jx}^2+\hat{C}_{jy}^2\rangle$, respectively. The 1st-inequality then becomes the one previously studied in \cite{Cavalcanti},  where the quadrature amplitudes or their higher-order versions were considered for the observables ${\hat C}_{jx}$ and ${\hat C}_{jy}$. However, it has been shown that the violation of such inequalities needs at least 5-mode entangled states with finite detector-efficiency requirement, which makes the experimental test rather demanding. On the other hand, the 2nd-inequality is a new one previously not known, which may deserve further detailed investigations elsewhere.

In summary, the violation of our inequalities as reported in \cite{Nha} provides a physically meaningful signature of nonlocality though not in the strongest sense due to the physical constraint used.
We also point out some interesting features that may come out of such physical
constraints, e.g., the interplay of wave- and particle-like properties in revealing nonlocal correlation and experimental advantages \cite{Nha}. Such a violation also clearly verifies quantum entanglement  (of negative-partial transpose character) as proved in \cite{Nha}, remarkably completely insensitive to detector efficiency. It can be readily shown that the violation is attributed to the quantum commutation rule, $[\hat{a},\hat{a}^\dag]=1$, which is the only component nonexistent in classical probabilistic descriptions; if the observables all commute with each other, our inequalities are readily satisfied. It may thus be used as a test of quantumness as also suggested by CS \cite{Daniel}.
\newline
*hyunchul.nha@qatar.tamu.edu

\end{document}